\documentclass[aps,prl,twocolumn,groupedaddress]{revtex4}
\usepackage[pdftex,colorlinks,bookmarks]{hyperref}
\usepackage{amsfonts}
\usepackage{amsmath}
\usepackage{amssymb}
\usepackage{graphicx}

\input{tcilatex}
\begin{document}

\title{Optimal approach to quantum communication using dynamic programming}
\author{Liang Jiang$^{1}$}
\author{Jacob M. Taylor$^{2}$}
\author{Navin Khaneja$^{3}$}
\author{Mikhail D. Lukin$^{1}$}
\affiliation{$^{1}$Department of Physics, Harvard University, Cambridge, Massachusetts
02138}
\affiliation{$^{2}$Department of Physics, Massachusetts Institute of Technology,
Cambridge, Massachusetts 02139}
\affiliation{$^{3}$School of Engineering and Applied Sciences, Harvard University,
Cambridge, Massachusetts 02138}
\date{\today }

\begin{abstract}
\end{abstract}

\maketitle

\textbf{Reliable preparation of entanglement between distant systems is an
outstanding problem in quantum information science and quantum
communication. In practice, this has to be accomplished via noisy channels
(such as optical fibers) that generally result in exponential attenuation of
quantum signals at large distances. A special class of quantum error
correction protocols---quantum repeater protocols---can be used to overcome
such losses. In this work, we introduce a method for systematically
optimizing existing protocols and developing new, more efficient protocols.
Our approach makes use of a dynamic programming-based searching algorithm,
the complexity of which scales only polynomially with the communication
distance, letting us efficiently determine near-optimal solutions. We find
significant improvements in both the speed and the final state fidelity for
preparing long distance entangled states.}

\section{Introduction}

Sequential decision making in probabilistic systems is a widely studied
subject in the field of economics, management science and engineering.
Applications range from problems in scheduling and asset management, to
control and estimation of dynamical systems ~\cite{Bertsekas2000}. In this
paper we make the first use of these techniques for solving a class of
decision making problems that arise in quantum information science~\cite%
{Nielsen2000,Bouwmeester2000}. Specifically we consider the optimal design
of a so-called quantum repeater for quantum communication. Such repeaters
have potential application in quantum communication protocols for
cryptography~\cite{Bennett1992,Bennett1993,Zukowski1993} and information
processing~\cite{Gottesman99}, where entangled quantum systems located at
distant locations are used as a fundamental resource. In principle this
entanglement can be generated by sending a pair of entangled photons through
optical fibers. However, in the presence of attenuation, the success
probability for preparing a distant entangled pair decreases exponentially
with distance~\cite{GisinRMP}.

Quantum repeaters can reduce such exponential scaling to polynomial scaling
with distance, and thus provide an avenue to long distance quantum
communication even with fiber attenuation. The underlying idea of quantum
repeater \cite{Briegel1998,Dur1999}\ is to \emph{generate} a backbone of
entangled pairs over much shorter distances, store them in a set of
distributed nodes, and perform a sequence of quantum operations which only
succeed with finite probability. \emph{Purification} operations \cite%
{Bennett1996a,Deutsch1996} improve the fidelity of the entanglement in the
backbone, while \emph{connection} operations join two shorter distance
entangled pairs of the backbone to form a single, longer distance entangled
pair. By relying on a quantum memory at each node to let different sections
of the repeater re-attempt failed operations independently, a high fidelity
entangled state between two remote quantum systems can be produced in
polynomial time. A quantum repeater \emph{protocol} is a set of rules that
determine the choice and ordering of operations based upon previous results.
An optimal protocol is one that produces entangled pairs of a desired
fidelity in minimum time within the physical constraints of a chosen
implementation.

The complexity of finding the optimal repeater protocols can be understood
via the following analogous example problem \cite{Bertsekas2000}: given a
sequence of rectangular matrices $M_1M_2 \dots M_n$, such that $M_k$ is $d_k
\times d_{k+1}$ dimensional, find the optimal order of multiplying the
matrices such that the number of scalar multiplications is minimized. This
is a typical example of a nesting problem, in which the order in which
operations are carried out effects the efficiency. For example, if $M_1 = 1
\times 10$, $M_2 = 10 \times 1$ and $M_3 = 1 \times 10$, then $(M_1 M_2)M_3$
takes only $20$ scalar operations, while $M_1(M_2 M_3)$, requires $200$
scalar multiplications. A brute force enumeration of all possible nesting
strategies and evaluation of their performance is exponential in $n$. To
solve this problem more efficiently, we observe that the optimal nesting
strategy $(M_1 \dots (\dots) \dots M_p)(M_{p+1} \dots M_n)$ should carry out
the solution to its subparts optimally, i.e., the nesting $(M_1 \dots
(\dots) \dots M_p)$ should represent the best nesting strategy for
multiplying $M_1M_2 \dots M_p$. This is the well-known dynamic programming
strategy \cite{Bertsekas2000}, in which one seeks to optimize a problem by
comparing different, already optimized sub-parts of the problem. Dynamic
programming enables us to find the optimal solution to the original problem
in time that is polynomial in $n$.

Quantum repeaters also have a nested (self-similar) structure, in which
shorter distance entanglement is used to create longer distance
entanglement, which is then used in turn for further extending the distance
between entangled pairs. This structure allows us to use the methods of
dynamic programming to find optimal nesting strategies for designing quantum
repeater protocols.

We now proceed to detail the specific optimization problem, then discuss our
dynamic programming solution to the problem. We next examine two
representative schemes that we wish to optimize [the Briegel \emph{et al.}
scheme (BDCZ scheme), Refs.~\cite{Briegel1998,Dur1999}, and the Childress
\emph{et al.} scheme (CTSL scheme), Refs.~\cite{Childress2005,Childress2006}%
], and find significant improvements in both preparation time and final
fidelity of long distance entangled pairs.

\section{Dynamic Programming Approach}

\paragraph{\textbf{General Quantum Repeater Protocol.}}

\begin{figure}[ptb]
\begin{center}
\includegraphics[height=7cm]{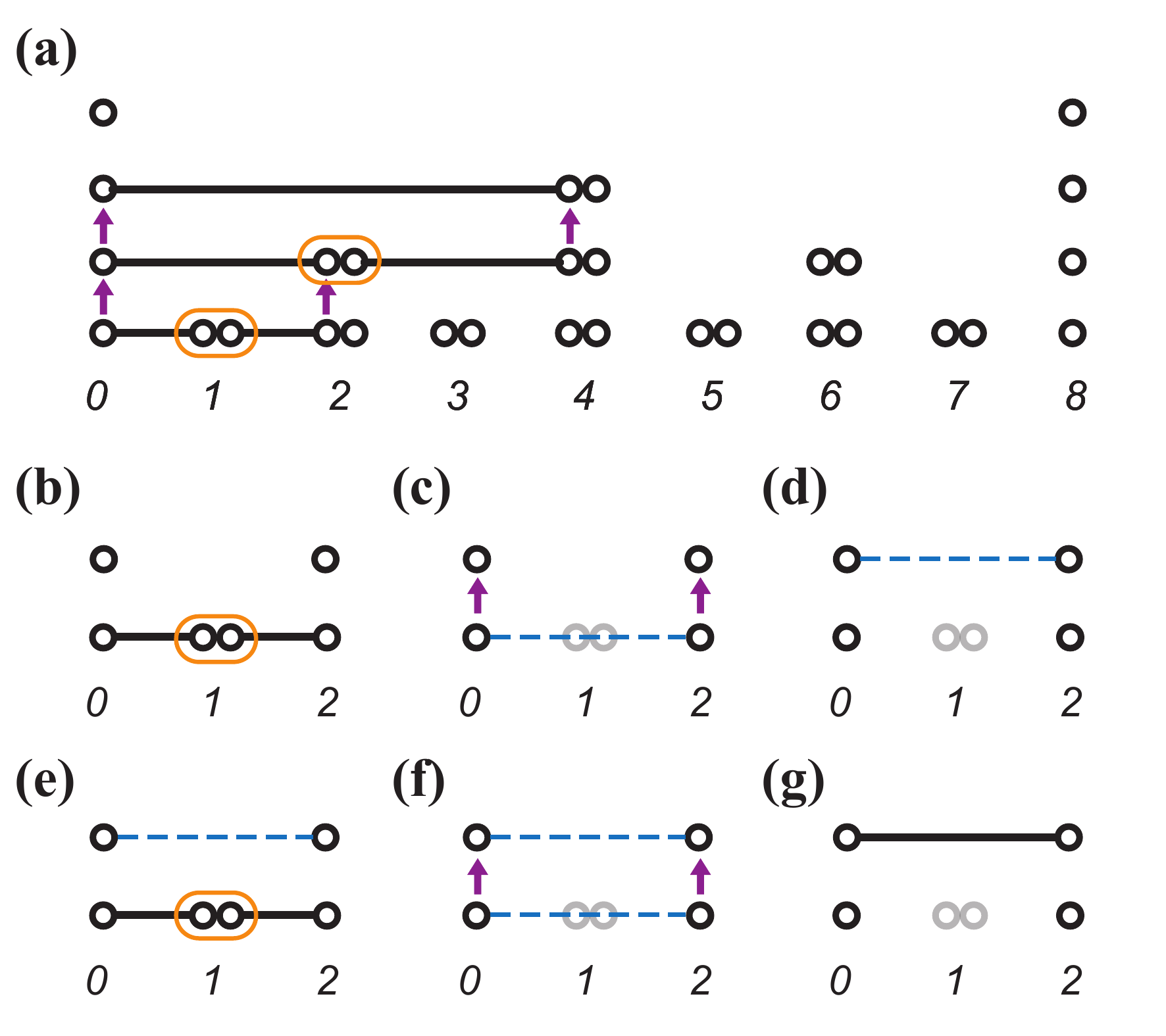}
\end{center}
\caption[figBDCZ]{\textbf{Quantum repeater scheme from Refs.~\protect\cite%
{Briegel1998,Dur1999} (BDCZ scheme).} \textbf{(a)} In a typical realization
with $N+1=9$ nodes, the number of qubits per nodes is bounded by $2\log_{2}2N
$ $=8$. \textbf{(b)-(d)} Two entangled pairs with distance $1$ are connected
(orange rounded rectangle) at node \#1 to produce an entangled state with
distance $2$, which is stored (purple arrows) in the qubits at higher level.
\textbf{(e)-(g)} Another entangled state with distance $2$ is produced to
purify (purple arrows) the entangled state stored in qubits at higher level.
Similarly, entangled states with distance $2^n$ can be connected to produce
entangled state with distance $2^{n+1}$, which may be further purified, as
indicated in \textbf{(a)}.}
\label{figBDCZ}
\end{figure}

\begin{figure}[ptb]
\begin{center}
\includegraphics[height=13.5 cm]{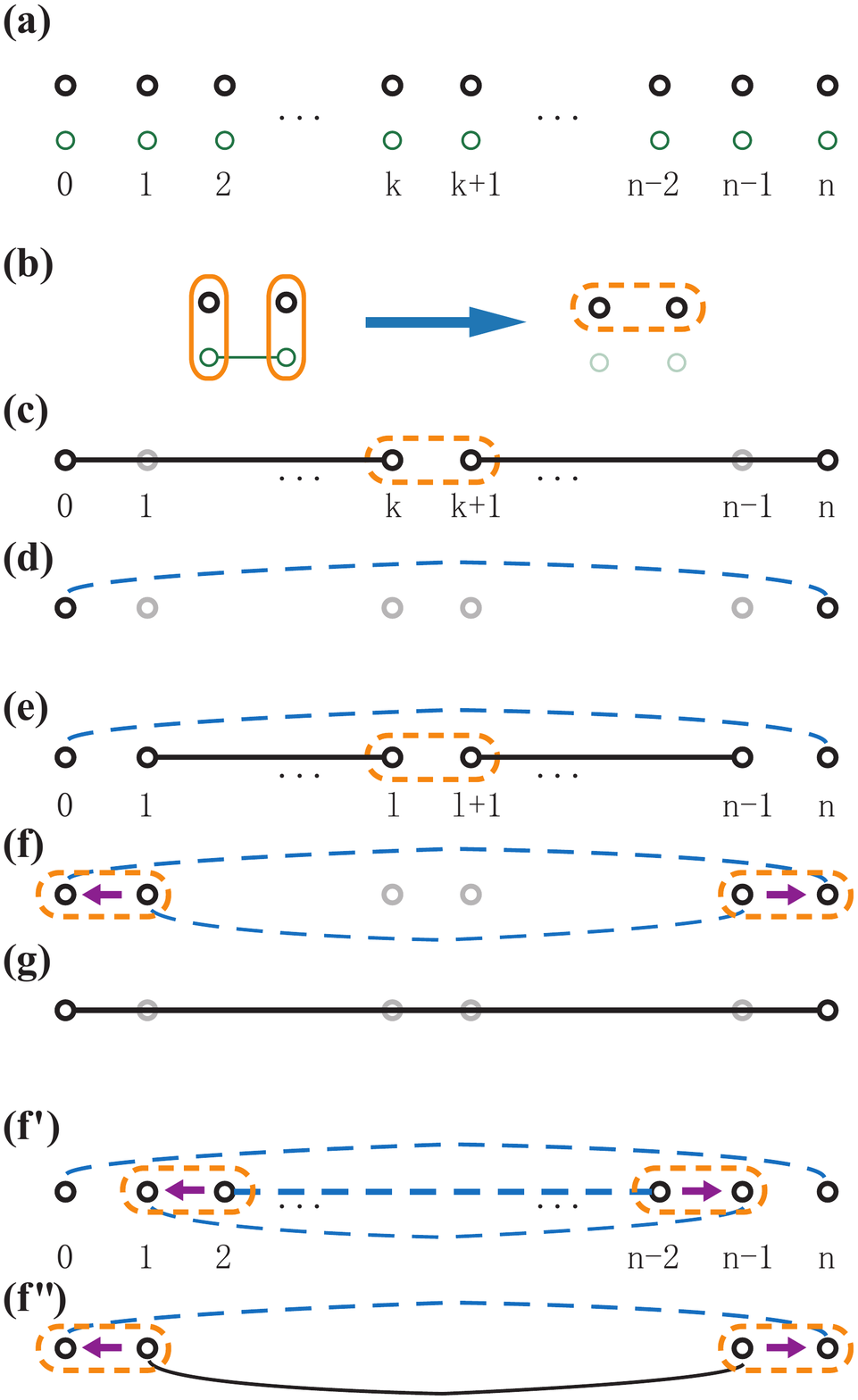}
\end{center}
\caption[figCTSL]{\textbf{Quantum repeater scheme from Refs.~\protect\cite%
{Childress2005,Childress2006} (CTSL scheme).} \textbf{(a)} This scheme has
exactly two qubits per node. The communication qubits (green nodes) are used
for entanglement generation and short-term storage; the storage qubits
(black nodes) are used for long-term storage. \textbf{(b)} With the help of
local gates (orange solid rounded rectangles) between communication and
storage qubits, the entangled state between communication qubits can be used
to implement entangling gates (e.g. the Controlled NOT gate) between storage
qubits from neighboring nodes. The effective remote gate is highlighted by
the orange dashed rounded rectangle. Such remote gate is sufficient for
entanglement connection and purification of storage qubits. The
communication qubits are omitted in the following plots. \textbf{(c)-(d)}
Entanglement connection to produce an unpurified entangled state with
distance $n$. \textbf{(e)-(g)} Entanglement connection to produce an
unpurified entangled state with distance $n-2$ to purify the entangled state
with distance $n$. \textbf{(f')(f\textquotedblright)} Illustration of
multi-level pumping. An entangled state with distance $n-4$ is used to
purify an entangled state with distance $n-2$, and the (purified) latter is
used to purify an entangled state with distance $n$. The only difference
between (f) and (f\textquotedblright) is the fidelity of the entangled state
with distance $n-2$. The latter has higher fidelity.}
\label{figCTSL}
\end{figure}

Quantum repeater protocols have a \emph{self-similar structure}, where the
underlying operations at each stage of the repeater have the same basic
algorithms. In other words, the structure of the problem remains the same at
each stage, while the parameters can be different. A generic quantum
repeater consists of three kinds of operations: entanglement generation,
entanglement connection, and entanglement purification. Entangled pairs are
first generated and stored over a short distance $L_{0}$. At the first
nesting level, two entangled pairs of distance $L_{0}$ can be extended to
distance $L_{1}\sim2L_{0}$ via entanglement connection \cite{Zukowski1993}.
Due to limited fidelity of the short pairs and the imperfections from the
connection operations, the fidelity of the longer pair produced by
connection is generally lower than those of the shorter ones. Nevertheless,
the fidelity of the longer pair can be improved via entanglement
purification, which is able to extract high-fidelity entangled pairs from an
ensemble of low-fidelity ones using operations that are local (restricted to
qubits within a given node)~\cite{Bennett1996a,Deutsch1996}. An efficient
approach of entanglement purification is entanglement pumping \cite%
{Briegel1998,Dur1999}, which purifies one and the same pair by using
low-fidelity pairs with constant fidelity. \footnote[1]{%
In principle, there exist repeater schemes (see \cite{Dur1999} and
references therein) that work much faster. For those schemes, however, the
number of memory qubits per repeater station scales at least linearly with
the final distance, which make them impractical.} Thus, at the $\left(
k+1\right) $th nesting level, the three underlying operations (preparation
at distance $L_{k}$, connection, and purification) lead to preparation at a
distance $L_{k+1}\sim2L_{k}$. \footnote[2]{%
Since we must wait for entanglement generation and purification to succeed
before proceeding to the next nesting level, the overall time for successful
pair generation is generally much longer than that of classical
communication over the given distance.}

\paragraph{\textbf{Inductive Optimization.}}

We now define the optimization problem:

\begin{description}
\item[Def:\ ] For given physical resources, desired distance $L_{\mathrm{%
final}}$, and final fidelity $F_{\mathrm{final}}$, an \emph{optimal protocol}
minimizes the expected time to have an entangled pair of fidelity $F \geq F_{%
\mathrm{final}}$ at a distance $L \geq L_{\mathrm{final}}$.
\end{description}

To solve this optimization problem, the choice of parameters for the quantum
operation cannot be viewed in isolation---one has to trade off the desire of
low present cost (in terms of time) with the undesirability of high future
costs. If one tries to enumerate and test all possible adjustable
parameters, the complexity to search for the optimized implementation scales
at least exponentially with the number of repeater nodes. A simple example
is provided if we make our only adjustable parameter choosing between zero
and one purification step at each stage of the protocol. For the BDCZ scheme
with $128+1$ repeater nodes, there are already $2^{128}\gtrsim10^{38}$
possibilities, which is beyond the capability of current computers. Thus a
systematic searching method is needed to find the optimized implementation
out of such a huge parameter space.

Based on the above self-similar structure, we may express the optimized
protocol to produce long entangled pairs in terms of a set of optimized
protocols for producing shorter pairs. The general searching procedure can
be performed inductively, as detailed in Table \ref{Table1}. We make a
discrete set of target fidelities (see Supporting Methods for details), $%
F=\{f_1,f_2,\ldots,f_q\}$, such that only a finite number of different
optimal protocols with shorter distances need to be developed. The
complexity for each step of our optimization procedure is shown in the
table; the full procedure scales as ${\mathcal{O}}(q^2 n^2)$, where the ${%
\mathcal{O}}(n)$ repetitions of step 3 take the most time. \footnote[3]{%
For the CTSL scheme, there will be another ${\mathcal{O}}(q)$ overhead,
associated with ${\mathcal{O}}(q)$ possible fidelity choices of the
entangled state for the Controlled NOT gate (Fig. \ref{figCTSL}b).} In
practice, we found the full search of step 3 to be unnecessary---the search
can be restricted to pairs of distance $n/2 \pm {\mathcal{O}}(\log(n)) $,
leading to complexity ${\mathcal{O}}(q^2 n \log(n))$.

\begin{table}[tbp]
\centering
\begin{tabular}{p{8.5cm}}
\hline
\multicolumn{1}{|p{8.5cm}|}{\textbf{1. \ }Find and store implementations
that optimize the average time (for all fidelities $f_1,\ldots,f_q$) with $%
dist=n=1$, taking ${\mathcal{O}}(q)$} \\ \hline
\multicolumn{1}{|p{8.5cm}|}{\textbf{2. \ }Assume known optimized
implementations (for all fidelities) with $dist\leq n$} \\ \hline
\multicolumn{1}{|p{8.5cm}|}{\textbf{3. \ }Find optimized implementations to
produce unpurified pairs (for all fidelities) with $dist=n+1$ by trying
(connecting) all combinations of known optimized implementations with $%
dist\leq n$, with complexity of order ${\mathcal{O}}(q^2 n)$} \\ \hline
\multicolumn{1}{|p{8.5cm}|}{\textbf{4. \ }Find optimized implementations to
produce purified pairs (for all fidelities) with $dist=n+1$\ by trying all
combinations of unpurified pairs with $dist=n+1$, pumping for $%
m=0,1,2,3,\cdots$\ times; complexity goes as ${\mathcal{O}}(m_{\mathrm{max}}
q^2)$} \\ \hline
\multicolumn{1}{|p{8.5cm}|}{\textbf{5. \ }Store the optimized
implementations (for all fidelities) with $dist=n+1$, based on step 4.} \\
\hline
\multicolumn{1}{|p{8.5cm}|}{\textbf{6. \ }Replace $n$\ by $n+1$, and go to
step 2.} \\ \hline
\end{tabular}%
\caption{Inductive search using dynamic programming }
\label{Table1}
\end{table}

\paragraph{\textbf{Repeater Schemes and Physical Parameters.}}

So far, we have only taken a general perspective in explaining quantum
repeater protocols and describing the procedure of inductive searching using
dynamic programming. In this subsection, we specify the parameters to be
optimized by examining the schemes of the quantum repeater, physical
restrictions on entanglement generation for current techniques, and the
error models of local quantum gates. Only with a functional relationship
between physically adjustable parameters and repeater operation outputs, can
we find the optimized implementations for procedure 1, 3 and 4 in Table~\ref%
{Table1}.

There are several different \emph{schemes} for building a quantum repeater
that differ primarily in the amount of physical resources utilized. For
example, in the BDCZ scheme \cite{Briegel1998,Dur1999} (Fig.~\ref{figBDCZ}),
the maximum number of qubits in the quantum memory (to store intermediate
states for connection and purification) required for each repeater node
increases logarithmically with the total number of repeater nodes. In the
CTSL scheme \cite{Childress2005,Childress2006} (Fig.~\ref{figCTSL}), an
efficient way to use quantum memory is proposed, and only \emph{two} qubits
are needed for each node, regardless of the total number of repeater nodes.
One of the two qubits is called the \emph{communication qubit}, which is
optically active and can be used to generate entanglement with other
communication qubits from neighboring nodes. The other qubit is called the
\emph{storage qubit}, which can be used to store quantum state over very
long time. As shown in Fig.~\ref{figCTSL}(b), with the help of local gates
(orange solid rounded rectangles) between communication and storage qubits,
the entangled state between communication qubits can be used to implement
teleportation-based gates (e.g., the Controlled NOT gate) between storage
qubits from neighboring nodes \cite{Gottesman99,JTSL}. Such remote gates
(orange dashed rounded rectangle) are sufficient for entanglement connection
and purification of the storage qubits; communication qubits are providing
the necessary resource mediating the gates between remote storage qubits.
For clarity, we will omit the communication qubits in the following
discussion, but still keep track of the mediated operation between remote
storage qubits.

To model errors in the physical operations, we need to introduce a number of
parameters determined by the quantum hardware. For entanglement generation,
the relationship between the fidelity of the elementary pair, $F_{0}$, and
the generation time, $\tau _{e}$, depends on the physical parameters (such
as the signal propagation speed, $c$, the fiber attenuation length, $L_{att}$%
, the efficiency of single photon collection and detection, $\varepsilon $,
and the distance of elementary pair, $L_{0}$) and the specific approach to
generate entanglement. For example, for the entanglement generation approach
using scattering as proposed in Refs.~\cite{Childress2005,Childress2006}, $%
F_{0}=F_{0}\left( \tau _{e}\right) =\frac{1}{2}\left\{ 1+\left[ 1-\frac{L_{0}%
}{\tau _{e}c}e^{L_{0}/L_{att}}\right] ^{2\left( 1-\varepsilon \right)
/\varepsilon }\right\} $.

For entanglement connection and pumping, the dominant imperfections are
errors from measurement and local two-qubit gate, which we model with a
depolarizing channel. In particular, the model for measurement is quantified
by a reliability parameter \cite{Briegel1998,Dur1999}, $\eta$, which is the
probability of faithful measurement. For example, a projective measurement
of state $\left\vert 0\right\rangle $ would be
\begin{equation*}
P_{0}=\eta\left\vert 0\right\rangle \left\langle 0\right\vert +\left(
1-\eta\right) \left\vert 1\right\rangle \left\langle 1\right\vert .
\end{equation*}
Similarly, the model for local two-qubit gate is characterized by a
reliability parameter \cite{Briegel1998,Dur1999}, $p$. With probability $p$,
the correct operation is performed; otherwise the state of these two qubits
is replaced by the identity matrix \cite{Briegel1998,Dur1999}. For example,
the action on a two qubit operation $U_{ij}$ would be
\begin{equation*}
U_{ij}\rho U_{ij}^{\dag}\rightarrow pU_{ij}\rho U_{ij}^{\dag}+\frac{1-p}{4}%
\text{Tr}_{ij}\left[ \rho\right] \otimes I_{ij},
\end{equation*}
where $Tr_{ij}\left[ \rho\right] $ is the partial trace over the subsystem $%
i $ and $j$, and $I_{ij}$ is the identity operator for subsystem $i$ and $j$%
. Generally, the reliability parameters ($\eta$ and $p$) should be
reasonably high (i.e., above some thresholds \cite{Briegel1998,Dur1999}), so
that the suppression of error from entanglement pumping dominates the new
errors introduced by entanglement connection and entanglement pumping.
\footnote[5]{%
We neglect the time associated with local operations, which is usually much
shorter than the communication time between neighboring repeater stations.
Non-negligible gate operation time can be easily included in our
optimization.}

\paragraph{\textbf{Optimization Parameters.}}

We now list the adjustable parameters we can optimize over during procedures
1, 3 and 4 in Table~\ref{Table1}.

\begin{enumerate}
\item During the entanglement generation, there is freedom to choose the
generation time $\tau_{e}$, which is determined by the success probability
and the communication time. Generally, the higher the success probability,
the shorter the generation time and the lower the fidelity of the entangled
state, so the generation time and the fidelity should be balanced \cite%
{Childress2005,Childress2006}.

\item During the entanglement connection, the distances of two shorter pairs
can be adjusted, while the total distance is kept unchanged.

\item During entanglement purification, the number of steps is also
adjustable, which should balance the gain in fidelity and the overhead in
time.
\end{enumerate}

\paragraph{\textbf{Additional Operations.}}

Besides the above operations from the original quantum repeater schemes,
there are some \emph{additional} operations that might be useful. For
example, we may skip several intermediate repeater nodes (\emph{node-skipping%
}) to generate entanglement between distant nodes directly with a
substantially lower success probability. Also, during entanglement pumping,
we might consider \emph{multi-level pumping} \cite{Dur2003}, which is to
nest several levels of entanglement pumping together before the next level
of entanglement connection (Fig.~\ref{figCTSL}(f\textquotedblright)).
Multi-level pumping can produce entangled pair with higher fidelity than
single-level pumping. Such additional operations can be easily incorporated
into the search procedures 1, 3, and 4 in Table~\ref{Table1}. We will show
that the dynamic programming approach can use these additional operations
appropriately, to reduce the average time, extend the upper bounds for
achievable final fidelity, and even improve the threshold for the
reliability parameters of $p$ and $\eta$.

\section{Results and Discussion}

\paragraph{\textbf{Improvement of BDCZ and CTSL Schemes.}}

\begin{figure}[t]
\begin{center}
\includegraphics[width=8.5cm]{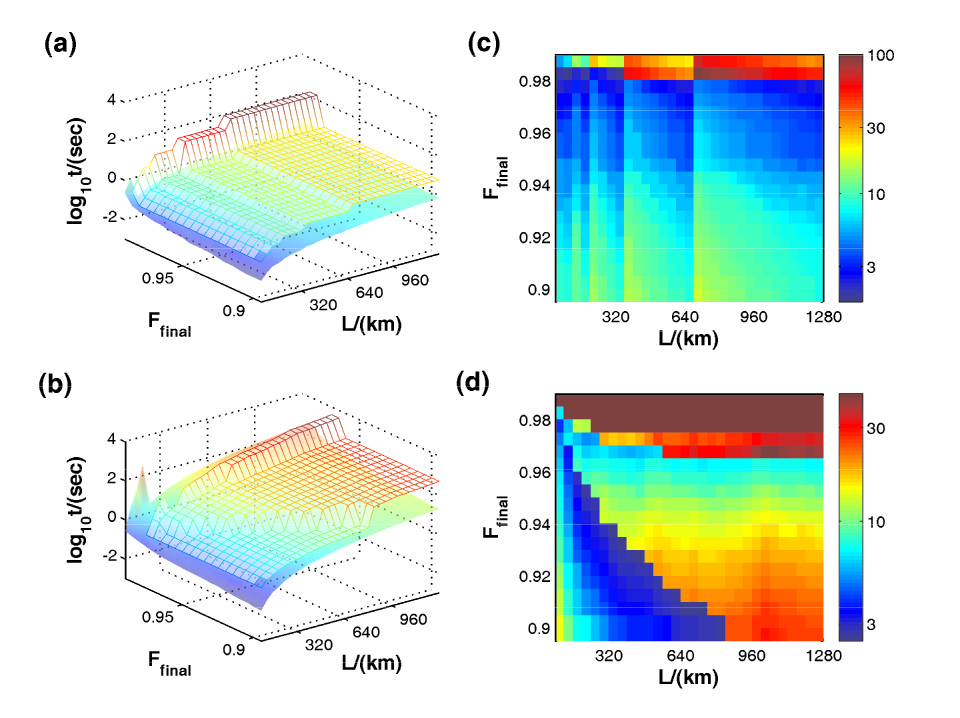}
\end{center}
\caption[figRatio]{\textbf{Plots of time profiles and improvement factors.}
Speed-up in time associated with various final distance and fidelity.
\textbf{(a)} $t\left( F_{final},L\right) $ for unoptimized (meshes) and
optimized (smooth surface) implementations of the BDCZ scheme; \textbf{(b)}
for the CTSL scheme. \textbf{(c)} Pseudocolor plot of the improvement
factor, $t_{unopt}/t_{opt}$, for the BDCZ scheme; \textbf{(d)} for the CTSL
scheme, in the region ($F_{final}>97.5$), the improvement factor $%
t_{unopt}/t_{opt}\rightarrow \infty$. The default parameters are $%
L_{att}=20km$, $\protect\varepsilon=0.2$, and $p=\protect\eta=0.995$.}
\label{figRatio}
\end{figure}

With procedure as listed in Table~\ref{Table1}, we implemented a computer
program to examine the mean time to prepare entangled pairs and to search
according to our dynamic programming prescription through the parameter
space outlined above. We looked for optimal protocols for a quantum repeater
for all distances $\leq 1280$ km and target fidelities $\geq 0.8$. Unless
otherwise specified, we use $L_{att}=20km$, $\varepsilon=0.2$, and $%
\eta=p=0.995$ for the rest of the discussion. We first fix $L_{0}=10km$, and
we will consider the optimization of $L_{0}$ to justify such choice later.
To visualize the results, the profile of the optimized time (smooth surface)
is plotted in Fig.~\ref{figRatio}(a)(b) with respect to the final distance
(from $10km$ to $1280km$) and the fidelity (from $0.90$ to $0.99$) for both
the BDCZ and the CTSL schemes. The calculation optimizes over the elementary
pair generation (both distance and generation time), the connecting
positions, and the number of pumping steps, with spacing between neighboring
repeater nodes of $10km$; both additional operations (node-skipping and
multi-level pumping) are also included for the optimization. For comparison,
the unoptimized time profiles (meshes) for the BDCZ and the CTSL schemes are
also plotted. The unoptimized protocol assumes fixed elementary pair
fidelity ($F_{0}=0.96$ and $0.99$ for BDCZ and CTSL, respectively), simple
connection patterns (detailed in Ref.~\cite{Briegel1998,Dur1999} and \cite%
{Childress2005,Childress2006}), and constant number of pumping steps.

As expected, the unoptimized protocol always takes longer time than the
optimized protocol for the same final distance and target fidelity. Time
profiles for the unoptimized protocols have \emph{stair-like jumps} in Fig.~%
\ref{figRatio}(a)(b). For the BDCZ scheme (Fig.~\ref{figRatio}(a)). The
jumps occurring with increasing distance (occurring at distances $%
L/L_{0}=2^{p}+1=1,3,5,9,17,33,\cdots$) are the results of time overhead from
the additional level of connection; the jump occurring with at $%
F_{final}=0.98$ is due to the sudden change in the number of pumping steps
from $1$ to $2$. Similarly, for the CTSL scheme (Fig.~\ref{figRatio}(b)),
the two jumps are due to the change of the number of pumping steps from $1$
to $2$ and finally to $3$. For the optimized protocols, the time increases
smoothly with increasing final distance and fidelity.

The improvement factor (i.e., the ratio between the times for unoptimized
and optimized protocols) is plotted for both the BDCZ and the CTSL schemes
in Fig.~\ref{figRatio}(c)(d). As we might expect, the previously mentioned
jumps lead to sharp \emph{stripes} where the improvement factor changes
discontinuously. There are several regions where the optimization gives
significant improvement. For example, for the BDCZ scheme, the vertical
bright stripes indicate that the optimization provides a time-efficient way
to generate entangled pairs for distance $\left( 2^{p}+\delta_{+}\right)
L_{0}$ (with $\delta_{+}>0$), gaining a factor of about $10$; the horizontal
bright stripes indicate that efficiently arranging the number of pumping
steps can also speed up the scheme by a factor of about $30$ or even more.
For most of the optimized protocols, a distant pair is divided into two
shorter pairs with similar distance and fidelity (symmetric partition), but
occasional asymmetric partitioning can further reduce the time by about $%
10\% $.

\begin{figure}[t]
\begin{center}
\includegraphics[width=8.5cm]{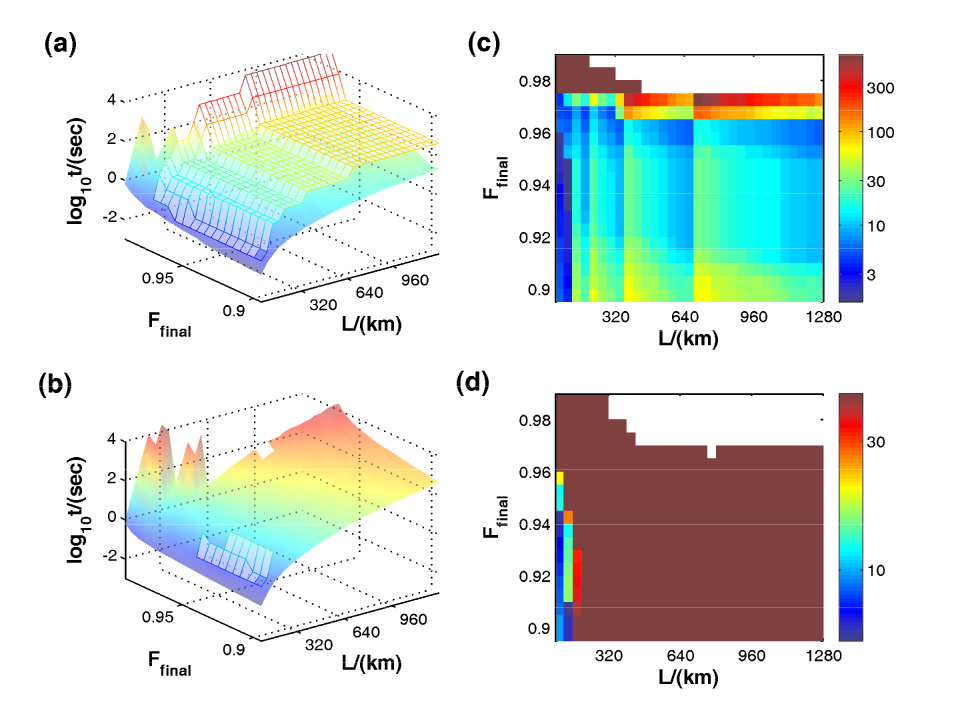}
\end{center}
\caption[figRatio2]{ \textbf{Plots of time profiles and improvement factors.}
The subplots are arranged in the same way as Fig.~\protect\ref{figRatio}.
Local operations have lower reliability parameters, $p=\protect\eta=0.990$. (%
\textbf{a})(\textbf{c}) For the BDCZ scheme, the optimization procedure only
improves the speed of the quantum repeater, and does not extend the
achievable region in the F-L plot. (\textbf{b})(\textbf{d}) For the CTSL
scheme, for distances longer than $200km$, the improvement factor, $%
t_{unopt}/t_{opt} \rightarrow \infty$. Here, the reliability parameter ($p=%
\protect\eta=0.990$) is \emph{insufficient} to create distant entangled
pairs with the unoptimized implementation, but the optimized implementation
(with multi-level pumping) is still able to create high-fidelity distant
entangled pairs, because multi-level pumping lowers the threshold of the
reliability parameters for the CTSL scheme. }
\label{figRatio2}
\end{figure}

For the BDCZ scheme, the correspondence between jumps and stripes
essentially accounts for all the features of the improvement plot (Fig.~\ref%
{figRatio}(c)). For CTSL scheme, however, besides the stripes, there is also
a region (with distance $L>100km$ and fidelity $F\gtrapprox97.5\%$) where
the improvement factor is infinity---optimization not only boosts the speed,
but also extends the upper bound of achievable fidelity for distant pairs.

\begin{figure*}[tb]
\begin{center}
\includegraphics[width=15 cm]{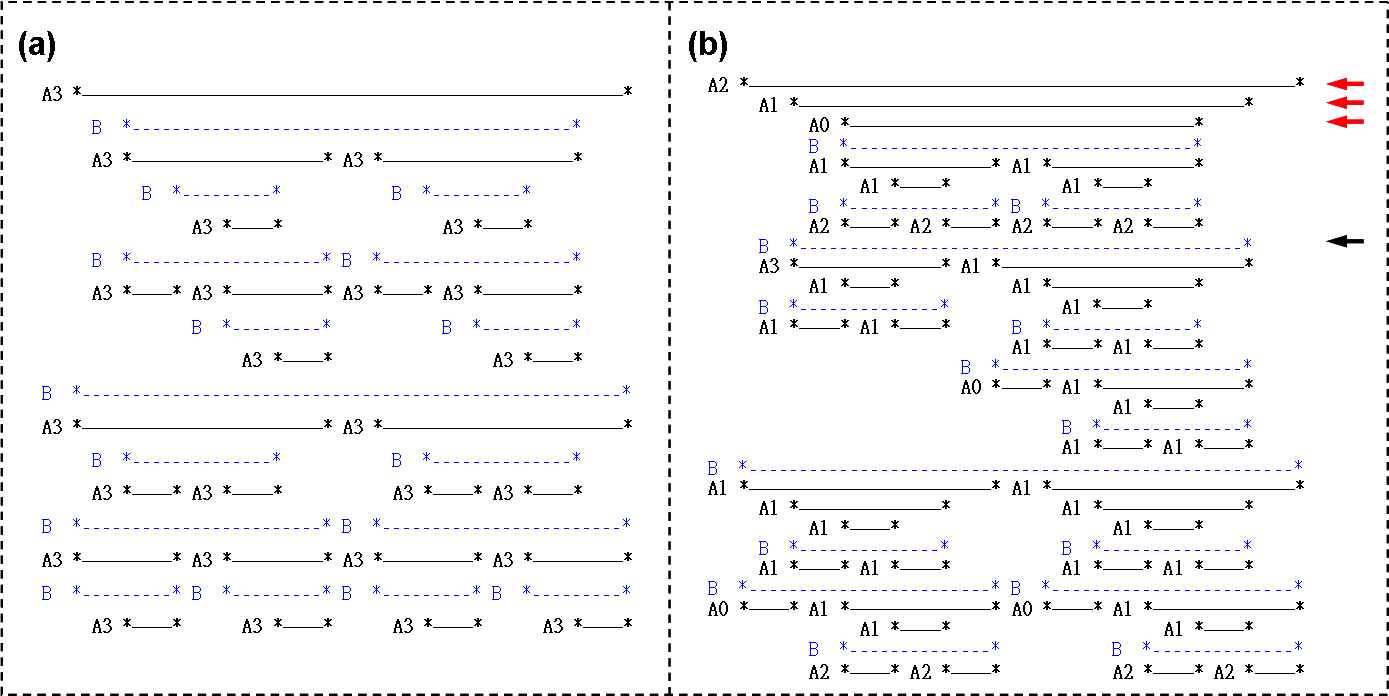}
\end{center}
\caption[figExample]{\textbf{Examples.} Two implementations with targeting
final distance $L=11L_{0}$ and fidelity $F_{final}=0.976$, using the CTSL
scheme. Each storage qubit is represented by "*". All the relevant entangled
states are shown. The order to produce these entangled states are from
bottom to top; states on the same row can be produced simultaneously. There
are two kinds of entangled states -- purified entangled states (type-A,
solid black line) and unpurified entangled states (type-B, dashed blue
line). On the left side of each purified entangled state, there is a label "A%
$k$", and this number $k$ indicates that this purified entangled states is
obtained from $k$ steps of entanglement pumping. \textbf{(a)} The
unoptimized (left) implementation has three pumping steps after each
entanglement connection, with average time of about $11\sec$ to produce the
pair wanted. \textbf{(b)} The optimized (right) implementation is from
optimization over pair generation time, connection position, and the number
of pumping steps. The optimized choice of connection position does not
necessarily break the long pair into two almost identical shorter pairs; for
example, the entangled state pointed by the black arrow in the $9$th row is
obtained by connecting two very different shorter pairs in the row below. In
addition, the possibility of multi-level pumping is also taken into account
during the dynamic programming. As pointed by the red arrows, the pair of
storage qubits in the third row pumps the pair in the second row, and the
latter pumps the pair in the first row. The average time is about $1.5\sec$
for the optimized implementation, about $8$ times faster than the
unoptimized one.}
\label{figExample}
\end{figure*}

We also study the improvement for other choices of reliability parameters, $%
p $ and $\eta$, especially those values close to the threshold \cite%
{Briegel1998,Dur1999}. Suppose the reliability parameters are $p=\eta=0.990$%
. In Fig.~\ref{figRatio2}(a)(c), we plot the speed-up in time associated
with various final distance and fidelity for the BDCZ scheme. For both
(optimized and unoptimized) protocols, the highest achievable fidelity is
approximately $97.5\%$ (compared to $99\%$ in Fig.~\ref{figRatio}(c)),
limited by errors from local operations. The improvement factor ranges
between $\left[ 1.5,600\right] $ (compared to $\left[ 1,100\right] $ in Fig.~%
\ref{figRatio}(c)). Apart from these differences, the key features
(horizontal and vertical stripes) of improvement from optimization are very
similar between Fig.~\ref{figRatio}(c) and Fig.~\ref{figRatio2}(c).

For the CTSL\ scheme, however, unoptimized and optimized protocols behave
very differently, when $p=\eta=0.990$. As shown in Fig.~\ref{figRatio2}%
(b)(d), the unoptimized protocol cannot effectively create entangled pairs
for distances longer than $200km$, while the optimized protocol is still
able to efficiently create distant entangled pairs with very high fidelity.
Thus our optimization lowers the threshold requirement for the CTSL scheme
of quantum repeater.

To understand the reason for the improvement of the highest achievable
fidelity (Fig.~\ref{figRatio}(d)) and the parameter threshold (Fig.~\ref%
{figRatio2}(d)), we examine the optimized protocol of CTSL scheme in the
next two subsections.

\paragraph{\textbf{Comparison between Optimized and Unoptimized Protocols.}}

We first compare the detailed procedures between two (optimized and
unoptimized) protocols using CTSL scheme to produce a pair with final
distance $L=11L_{0}$ and fidelity $F_{final}=97.6\%$, with default
reliability parameters $p=\eta=0.995$. We choose the highest fidelity
achievable by the unoptimized protocol, so that we will see almost all
features that give improvements. The results for the unoptimized protocol
(Fig.~\ref{figExample}(a)) follows Refs.~\cite{Childress2005,Childress2006}
exactly, while the optimized one (Fig.~\ref{figExample}(b)) is from our
systematic search using dynamic programming. They differ in the following
aspects: (1) during entanglement generation, the optimized implementation
generates elementary pairs with fidelity lower than $0.99$ to reduce the
generation time, and uses entanglement pumping afterwards to compensate the
fidelity loss; (2) during entanglement connection, the rule of producing
long pair from two almost identical shorter pairs is slightly modified
(e.g., the pair pointed by the black arrow in the $9$th row is from
connection of two quite different pairs in the $10$th row); (3) the number
of pumping steps after each connection varies from $0$ to $3$ for optimized
implementation; (4) finally, the optimized implementation uses multi-level
pumping, which will be discussed in detail in the next subsection. For
clarity, the additional operation of node-skipping is suppressed in the
optimization here. The overall average time is reduced from $11\sec$ to $%
1.5\sec$, improved by a factor of $8$.

\begin{figure}[ptb]
\begin{center}
\includegraphics[width=8.5cm]{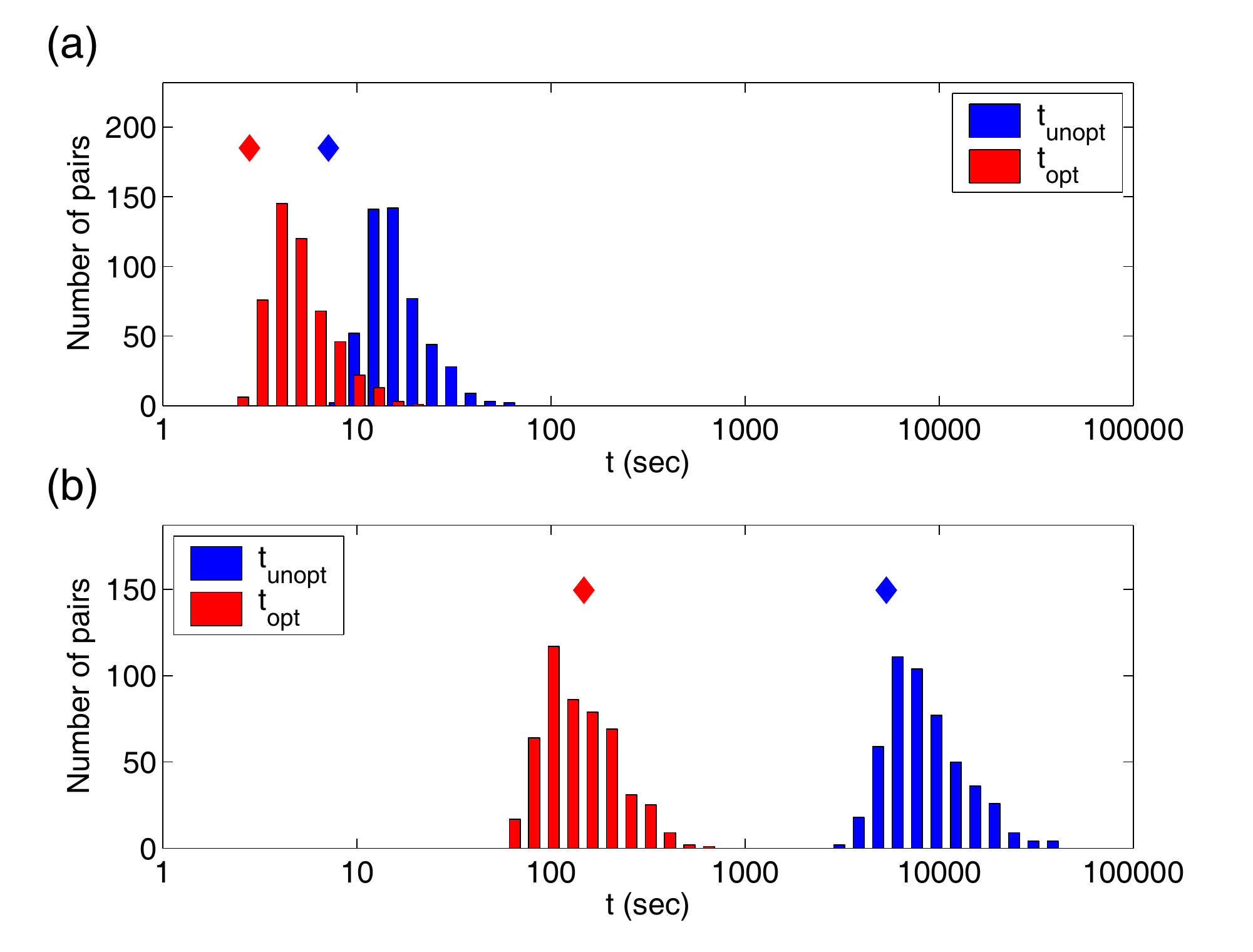}
\end{center}
\caption[figMC]{\textbf{Results of Monte Carlo simulation.} Monte Carlo
simulation for unoptimized/optimized implementations for \textbf{(a)} the
BDCZ scheme and \textbf{(b)} the CTSL scheme, with final distance $1280km$
and fidelity $0.97$. The time distributions for distant pairs are plotted,
with red (blue) bars for optimized (unoptimized) implementation. In each
plot, the red (blue) diamond indicates the estimated time from average-time
approximation for optimized (unoptimized) implementation. The average-time
approximation provides a good estimate up to some overall factor ($2\sim3$)
which is not very sensitive to the implementation.}
\label{figMC}
\end{figure}

Note that our optimization results based on average time approximation (see
Fig.~\ref{figMC} and Supporting Methods) are confirmed by the Monte-Carlo
simulation of the optimized protocols, verifying the substantial speed-up.

\paragraph{\textbf{Multi-Level Pumping.}}

We now consider the additional operation of multi-level pumping in more
detail. We discuss multi-level pumping only for the CTSL scheme, but not for
the BDCZ scheme. (In the BDCZ scheme, to introduce multi-level pumping
requires additional quantum memory qubits.) In the original unoptimized
protocol \cite{Childress2005,Childress2006}, the purified entangled state
with distance $n$ (between the $0$th and the $n$th nodes ($n>5$)) is
produced by entanglement pumping, and the entangled states used for pumping
(called pumping-pairs) are \emph{unpurified} entangled states with distance $%
n-2$ (Fig.~\ref{figCTSL}(f)). The fidelity of these pumping-pairs with
distance $n-2$ are limited by the connection operation, which imposes an
upper-bound for the fidelity of the purified pair with distance $n$. The
underlying restriction is that the pumping-pair is unpurified.

We may lift this restriction by allowing the use of a purified pumping pair.
This is multi-level pumping. For example, the pumping-pair with distance $%
n-2 $ may also be produced by entanglement pumping from pumping-pairs with
distance $n-4$ (Fig.~\ref{figCTSL}(f')), and so on. By doing multi-level
pumping, the fidelity of the pumping-pair with distance $n-2$ is increased
(Fig.~\ref{figCTSL}(f\textquotedblright)), and the same for the fidelity
upper-bound for the entangled state with distance $n$. While multi-level
pumping can increase the fidelity, it also slows down the repeater scheme.

When the reliability of local operations is above the threshold for the
unoptimized protocol (e.g. $p=\eta=0.995$), we find that multi-level pumping
is necessary only for the last two or three levels to the high-fidelity pair
we want to produce. Such multi-level pumping can be identified in the
optimized implementation -- for example, as indicated by red arrows\ in Fig.~%
\ref{figExample}(b), the pair of storage qubits in the third row pumps the
pair in the second row, and the latter pumps the pair in the first row. On
the other hand, when the reliability of local operations is below such
threshold (e.g. $p=\eta=0.990$), multi-level pumping is needed almost after
every entanglement connection.

If we exclude the possibility of multi-level pumping in dynamic programming,
the infinite improvement factor for pairs with distance $L>100km$ and
fidelity $F\gtrapprox97.5\%$ in Fig.~\ref{figRatio}(d) would disappear.
Similarly, in Fig.~\ref{figRatio2}(d), without multi-level pumping, there
would be no improvement of the parameter threshold, and even the optimized
protocol could not efficiently create distant ($L>200km$) entangled pairs.
For the CTSL scheme, multi-level pumping not only enables us to prepare
entangled pairs with very high fidelity, but also lowers the required
threshold of the reliability parameters ($p$ and $\eta$) for local
operations. Therefore, the flexibility to include additional operations in
our dynamic programming provides a new perspective on the optimization of
quantum repeater schemes.

\paragraph{\textbf{Other Improvements.}}

In addition to the previously discussed features in the plots of improvement
factor, there is an overall improvement for all final distances and
fidelities. Such overall improvement comes from the optimized choice of the
distance (by node-skipping) and the generation time for \emph{each}
elementary pair used. Such overall improvement is about $1.5$ (or $2\sim3$)
for the BDCZ (or CTSL) scheme, which indicates that the original choice of
uniform distance $L_{0}=10km$ and initial fidelities $F_{0}=96\%$ (or $99\%$%
) are quite close to the optimal.

Finally, we consider if it is possible to gain some additional speed-up if
we are allowed to choose the location of the nodes of the quantum repeater.
In order to answer this question, we discretize the distance into smaller
units, e.g. $1km\ll L_{att}$. Since the distance of each elementary pair is
determined by the dynamic programming, the optimized location of the nodes
can be inferred from the distances of the elementary pairs. We find that the
speed-up due to optimization over the location of the nodes is fairly small,
no more than $15\%$ in time (for cases with final distances larger than $%
200km$). Generally we find that as long as the node spacing is less than the
attenuation length ($L_{0}<L_{att}$), a quantum repeater can be implemented
almost optimally.

\paragraph{\textbf{Experimental Implications.}}

Throughout our analysis we have assumed relatively high fidelity of local
measurements and operations ($\eta=p=0.995$ or $0.99$) and memory times
exceeding total communication times. Recent experiments with tapped ions
\cite{Leibfried2003,Hume2007}, neutral atoms \cite{Beugnon2006}, and solid
state qubits \cite{Dutt2007} are already approaching these values of
fidelity and memory times. At the same time, high initial entanglement
fidelity ($F_{0}\approx96\%$ or $99\%$) is also needed for the optimized
protocols. Entanglement fidelity of about $90\%$ can be inferred from recent
experiments with two ions in independent traps \cite{Maunz2007}. While
optimization procedure can yield protocols compatible with fairly low
initial fidelity and high local error rates, in practice these errors
introduce a large overhead in communication time.

Besides the schemes considered here, there exist other quantum repeater
schemes, in particular the Duan \emph{et al.} scheme (DLCZ scheme) \cite%
{DLCZ2001} that requires a smaller set of quantum operation and relatively
modest physical resources. The original DLCZ scheme does not use active
entanglement purification and hence cannot correct arbitrary errors. In such
a case, optimization is straightforward and has been discussed in \cite%
{DLCZ2001}. Recently, the DLCZ\ scheme has been extended to include active
entanglement purification in order to suppress e.g. phase noises \cite%
{Jiang2007,Zhao2007}. The extended DLCZ scheme becomes very similar to the
BDCZ scheme in terms of the self-similar structure. The technique of dynamic
programming can be applied to optimize the extended DLCZ scheme as well.

\section{Conclusion and Outlook}

We have demonstrated how dynamic programming can be a powerful tool for
systematically studying the optimization of quantum repeater protocols. We
find substantial improvements to two specific repeater schemes~\cite%
{Briegel1998,Dur1999,Childress2005,Childress2006}. Beyond searching for
optimal choices in previously known elements of the schemes (entanglement
generation, connection, and pumping), our systematic study can also
incorporate more sophisticated additional operations, such as node-skipping,
multi-level pumping, and the flexible location of repeater stations. In
particular, our multi-level pumping procedure extends the maximum achievable
fidelity for distant pairs. It should be possible to include additional
possibilities to the optimization problem of quantum repeater, such as
different choices of entanglement generation and possibly more efficient
usage of local qubits~\cite{vanLoock2006,Ladd2006}. It would also be
interesting to study the optimization problem of quantum repeater with
finite storage time of the quantum memory~\cite{Hartmann2007,Collins2007}.
Even the optimized protocols have a rather limited speed (corresponding to
generation of one high-fidelity pair over $1280$ km in $1\sim 100$ s (see
Fig.~\ref{figMC}). Therefore, improvement of experimental techniques (to
obtain higher local operation fidelities and more efficient atom-photon
coupling) as well as development of new theoretical approaches to speed-up
quantum repeaters still remain an outstanding goal. Furthermore, the dynamic
programming techniques may find application in other outstanding problems in
quantum information science, such as the optimization of quantum error
correction for fault tolerant quantum computation. In particular, the
optimization of the network-based quantum computation scheme with minimal
resources~\cite{JTSL} might be possible.

\begin{acknowledgments}
The authors wish to thank Lily Childress and Wolfgang Dur for stimulating
discussions. This work is supported by NSF, ARO-MURI, ONR-MURI, the Packard
Foundations, and the Pappalardo Fellowship. N.K. acknowledges research
support from AFOSR FA9550-05-1-0443, NSF 0133673 and Humboldt foundation.
\end{acknowledgments}

\pagebreak

\section{Supporting Methods: Shape Parameter Approximation and Average Time
Approximation}

We use two important approximations throughout the analysis: the \emph{shape
parameter approximation} and \emph{average time approximation}.

In the shape parameter approximation \cite%
{Dur1999,Childress2005,Childress2006}, we use two numbers, the fidelity and
the shape parameter $\left( F,v\right) $ to classify non-ideal entangled
states during our inductive optimization procedure. For a pair of entangled
qubits, we assume that the density matrix is diagonalized in the Bell basis $%
\rho =diag\left( F_{1},F_{2},F_{3},F_{4}\right) $, with $F_{1}\geq F_{2}\geq
F_{3}\geq F_{4}$ by an appropriate ordering of the Bell basis. The fidelity $%
F\equiv F_{1}$ is the probability of finding the pair in the desired Bell
state. The shape parameter $v\equiv \frac{1}{2}\frac{F_{3}+F_{4}}{\left(
F_{2}+F_{3}+F_{4}\right) }$ measures, e.g., the relative ratio of bit error
to phase error in the generated pair. We may classify various entangled
states according to $\left( F,v\right) $, and use this classification to
facilitate the bookkeeping of different states. For intermediate distances,
we only keep track of the minimum time (and the associated density matrix)
for each class of states labeled by $\left( F,v\right) $, rather than for
each possible state. This significantly alleviates the computational storage
requirements of dynamic programming. Meanwhile, since we only use a subset
of states to optimize longer distance pairs, the obtained protocol might be
a little slower (no more than $10\%$) than the optimal one.

In our average time approximation, we only keep track of the average time
for entanglement generation, connection and pumping, instead of the full
distribution function. This underestimates the time for entanglement
connection, because the average time to generate both sub-pairs is longer
than the maximum of the average individual times for two sub-pairs.
Fortunately, our comparison between the average time approach and the Monte
Carlo simulations show that the average times from the average-time
approximation and from Monte Carlo simulations only differ by a factor of
about $2$ for both the BDCZ and the CTSL schemes. In Fig.~\ref{figMC}, the
time distributions from Monte Carlo simulation are plotted. %
\end{document}